# Estimation of coupling between oscillators from short time series via phase dynamics modeling: limitations and application to EEG data

*D.A. Smirnov[a], M.B. Bodrov[b], J.L. Perez Velazquez[c], R.A. Wennberg[d], and B.P. Bezruchko[a,b]*

[a]Saratov Branch, Institute of RadioEngineering and Electronics, Russian Academy of Sciences, Zelyonaya street 38, Saratov, 410019, Russia; e-mail: smirnovda@info.sgu.ru

[b]Saratov State University, Astrakhanskaya street 83, Saratov, 410012, Russia

[c]Hospital for Sick Children, Brain and Behaviour Programme and Division of Neurology, University of Toronto, 555 University Avenue, Toronto, Canada M5G 1X8

[d]Toronto Western Hospital, Krembil Neuroscience Centre, University of Toronto, 399 Bathurst Street, Toronto, Canada M5T 2S8

We demonstrate in numerical experiments that estimators of strength and directionality of coupling between oscillators based on modeling of their phase dynamics [D.A. Smirnov and B.P. Bezruchko, Phys. Rev. E **68**, 046209 (2003)] are widely applicable. Namely, although the expressions for the estimators and their confidence bands are derived for linear uncoupled oscillators under the influence of independent sources of Gaussian white noise, they turn out to allow reliable characterization of coupling from relatively short time series for different properties of noise, significant phase nonlinearity of the oscillators, and non-vanishing coupling between them. We apply the estimators to analyze a two-channel human intracranial epileptic electroencephalogram (EEG) recording with the purpose of epileptic focus localization.

An interdisciplinary problem of detecting interaction between oscillatory systems solely from their time realizations has attracted attention of researchers for a long time. Several approaches to its solution have been suggested within the framework of linear time series analysis and information theory. The most well-known of them are cross-correlation function, coherence function, and mutual information function, which are typically capable of detecting only the presence of interdependence. To detect coupling directionality, their generalizations exist, such as Granger causality[2], Geweke's spectra[3], and similar information-theoretic concepts[4,5]. Recently, there have been developed new approaches in nonlinear dynamics to reveal the presence of nonlinear interaction and its directionality. These nonlinear techniques are based either on analysis in state spaces[6-13] or investigation of phase dynamics[14-20]. The latter set of approaches includes an evolution map approach, based on modeling phase dynamics of the systems[17,18], and its extension for the case of relatively short time series[1]. The latter technique is shown to be often more sensitive to weak coupling than state space approaches, especially for the practically important case of short signals[21]. In addition, expressions for the confidence bands have been developed for the coupling estimators of Ref. 1 that increases reliability of the results. But all the formulas are rigorously valid only for weakly nonlinear and weakly coupled phase oscillators under the influence of independent sources of Gaussian white noise. In the present paper, we investigate practical limits of applicability of these formulas and show in numerical experiments that they are quite wide. Finally, application of the estimators to an intracranial EEG recording from an epileptic patient is presented.

## 1. INTRODUCTION

Characterization of coupling between two oscillatory systems from their time series is an important task in different fields of scientific research and practice, including climatology[22], electronics[23], and physiology[24]. Thus, a great deal of attention is paid nowadays to investigation of interaction between human cardio-vascular and respiratory systems[16,18,19,25-28] and to analysis of multichannel EEG and MEG recordings[9-13,15,29-35], in particular, with the purpose to localize epileptic foci[9,15,29,30,32,34]. Most of the well-known approaches, such as cross-spectral analysis and information-theoretic characteristics, are often insufficient to detect directional coupling from complex real-world signals. In the last years, new promising techniques are suggested by nonlinear dynamics, see comparative study of several approaches in Refs. 21 and 31.

One family of nonlinear approaches exploits the idea to analyze interdependencies between *phases* of the oscillatory systems. The most sensitive approach within this family involves construction of an empiric model for the phase dynamics and calculation of interaction strength from the values of its parameters. The idea is suggested originally in Ref. 17 and the technique to realize it is called "evolution map approach" (EMA). It is efficient for analysis of oscillatory processes unsynchronized with each other and exhibiting pronounced main rhythms of oscillations that allows to introduce well-defined phases. In its initial version, EMA provides reliable results for stationary time series of quite a considerable length, such as 5000 characteristic periods under moderate noise levels. A very similar approach is proposed by Kiemel et al[14,20].

However, in practice one often encounters *nonstationary* signals, e.g., EEG recordings are well-known to be highly nonstationary[36]. Thus, the problem of coupling characterization from short time series segments inevitably arises. To address it, special corrections have been introduced into formulas for the EMA coupling estimators, so that the latter become unbiased even in the case of relatively short time series (down to 50 basic periods), and expressions for their confidence bands have been derived in Ref. 1. The modified expressions for the coupling estimators are derived under the assumptions of linear uncoupled phase oscillators influenced by independent sources of Gaussian white noise. Their applicability in other cases has neither been rigorously proven, nor thoroughly investigated experimentally. Our purpose here consists in a systematical investigation of the limits of applicability of the modified EMA estimators. Relevance and applied importance of such a work is justified by a variety of situations, where one needs to detect weak coupling from short time series and the modified EMA appears very sensitive and reliable. Yet, under some conditions its efficiency deteriorates, so that other techniques can be more effective as discussed in Sec. 4.

The paper is organized as follows. We describe the modified EMA in Sec. 2.1 and a technique to find out limits of its applicability in Sec. 2.2. Results of investigation are reported in Sec. 3 where we show the influence of noise properties (Sec.3.1), individual nonlinearity of oscillators (Sec.3.2), coupling intensity (Sec.3.3), several factors together including the case of common source of the noise (Sec.3.4), and illustrate an application of the method with analysis of an epileptic EEG recording (Sec.3.5). Discussion of the relationships between the modified EMA and other approaches and summary of our results is presented in Sec.4. Too cumbersome formulas for the coupling estimators are given in Appendix.

## 2. METHODS

### 2.1. Modified evolution map approach

The main idea of the original method is to estimate how strongly future evolution of the phase of one system depends on the current value of the phase of the other system. To achieve this, one obtains time series of the oscillations' phases $\{\phi_1(t_1),...,\phi_1(t_N)\}$ and $\{\phi_2(t_1),...,\phi_2(t_N)\}$ from original time series of the two systems $\{x_1(t_1),...,x_1(t_N)\}$ and $\{x_2(t_1),...,x_2(t_N)\}$, $t_i = i\Delta t$, $\Delta t$ is sampling interval. An analytic signal is constructed for this purpose typically in one of the two ways. The most traditional one is to calculate Hilbert transform $y_k(t)$ of the observed signal $x_k(t)$:

$$y_k(t) = p.v. \int_{-\infty}^{\infty} \frac{x_k(t')dt'}{\pi(t-t')} \tag{1}$$

where *p.v.* stands for the Cauchy principal value. Then, one defines complex analytic signal $z_k(t) = x_k(t) + jy_k(t)$ [16,37]. The second approach is to define $z_k(t)$ via complex wavelet transform:

$$z_k(t) = \int_{-\infty}^{\infty} x_k(t')\psi\left(\frac{t'-t}{s}\right)dt', \tag{2}$$

where $\psi(t)$ is a complex wavelet function, $s$ its time scale. As a rule, Morlet wavelet $\psi(t) = \pi^{-1/4} \exp(j\omega_0 t)\exp(-t^2/2)$ is employed[38]. For any of these approaches, one defines an unwrapped phase $\phi_k(t)$ as the argument of the signal $z_k(t) = a_k(t)\exp(j\phi_k(t))$ increased by $2\pi$ after each complete revolution of the vector $z_k(t)$ about the origin[16]. Both approaches are closely related as shown in Ref. 39: the use of the complex wavelet transform corresponds to band-pass filtering of the signal $x_k(t)$ around the circular frequency $\omega_0/s$ with bandwidth determined by $\omega_0$ and subsequent calculation of the Hilbert transform to define the phase of the filtered signal. For any approach, sampling frequency for the original time series is desirable to be not less than 20 points per basic period to extract the phase without significant distortions[16,40].

After calculation of the phases, one constructs a mathematical model from their time realizations. Model structure is chosen based on the following considerations. In variety of situations, the phase dynamics of oscillators exhibiting a pronounced main rhythm are adequately described with stochastic differential equations of the form[41]

$$d\phi_{1,2}/dt = \omega_{1,2} + G_{1,2}(\phi_1,\phi_2) + \xi_{1,2}(t), \qquad (3)$$

where parameters $\omega_{1,2}$ govern oscillators' frequencies, $\xi_i(t)$ are independent Gaussian white noises with zero mean and autocorrelation functions (ACF) $\langle \xi_i(t)\xi_i(t')\rangle = \sigma_i^2 \delta(t-t')$. When dealing with discrete time series, it is convenient to consider a difference form of these equations

$$\Delta_{1,2}(t) = F_{1,2}[\phi_1(t),\phi_2(t),\mathbf{a}_{1,2}] + \varepsilon_{1,2}(t), \qquad (4)$$

where $\Delta_i(t) \equiv \phi_i(t+\tau) - \phi_i(t)$ are phase increments over fixed time interval $\tau$, $\varepsilon_i(t)$ zero-mean noises, $F_i$ trigonometric polynomials, $\mathbf{a}_i$ vectors of their coefficients. To construct a model (4), one specifies the orders of the polynomials $F_i$ and the interval $\tau$ which is usually equal to the basic period of oscillations[16]. Using the time series of phases, one gets estimates $\hat{\mathbf{a}}_i$ of the coefficients $\mathbf{a}_i$ via the least-squares routine. Then, one calculates strengths of influence of oscillators on each other from the model coefficients as explained below.

If the "true" equations for phase dynamics were known *a priori*, then the intensity $c_1$ of the influence of the second system on the first one (2→1) would be defined as the steepness of the dependence $F_1(\phi_2)$, and everything is the same for the intensity $c_2$ of the influence 1→2:

$$c_{1,2}^2 = \frac{1}{2\pi^2}\int_0^{2\pi}\!\!\int_0^{2\pi} [\partial F_{1,2}(\phi_1,\phi_2,\mathbf{a}_{1,2})/\partial \phi_{2,1}]^2 d\phi_1 d\phi_2 . \qquad (5)$$

Directionality index is determined by the difference between $c_1$ and $c_2$. It would be *true* coupling characteristics. However, one has only estimates of the coefficients $\hat{\mathbf{a}}_i$ obtained from a time series and needs to calculate estimates of $c_1$ and $c_2$ based on $\hat{\mathbf{a}}_i$. The most direct way is to use Eq. (5) substituting the estimates $\hat{\mathbf{a}}_i$ for the true values $\mathbf{a}_i$. But such estimators $\hat{c}_{1,2}$ appear "good" only for very long stationary signals whose length should be about 5000 basic periods for the sampling frequency 10-20 points per a basic period and moderate noise level[1,17,23]. For shorter time series often encountered in practice, these estimators turn out to be biased. The modified estimators $\hat{\gamma}_{1,2}$ for $c_{1,2}^2$ and the estimator $\hat{\delta} \equiv \hat{\gamma}_2 - \hat{\gamma}_1$ for the directionality index $\delta = c_2^2 - c_1^2$ are suggested in Ref. 1, see Appendix. Expressions for their 95% confidence bands are derived in the from $[\hat{\gamma}_i - 1.6\hat{\sigma}_{\hat{\gamma}_i}, \hat{\gamma}_i + 1.8\hat{\sigma}_{\hat{\gamma}_i}]$ and $\hat{\delta} \pm 1.6\hat{\sigma}_{\hat{\delta}}$ where $\hat{\sigma}_{\hat{\gamma}_i}$ and $\hat{\sigma}_{\hat{\delta}}$ are calculated from the *same* short time series. Under the assumption of linear uncoupled phase oscillators and independent sources of Gaussian white noise, these modified estimators are unbiased and provide the rate of erroneous conclusions about coupling presence and directionality less than 5 % for time series whose length may be as small as 50 basic periods.

2.2. Technique for investigation of applicability limits in numerical experiments

Expressions for the estimators $\hat{\gamma}_{1,2}$ and $\hat{\delta}$ are derived analytically for the system (3) with $G_{1,2} \equiv 0$ whose equations can be rewritten rigorously in the form[1]

$$\Delta_i(t) = \omega_{1,2}\tau + \varepsilon_i(t), \quad i = 1,2, \qquad (6)$$

where $\varepsilon_i$ are independent Gaussian noises with variances $\sigma_i^2 \tau$, their ACFs are linearly decreasing from $\sigma_i^2 \tau$ down to zero over the interval $[0, \tau]$. If one of the mentioned properties of the system (Gaussianity and independence of $\varepsilon_i$, forms of their ACFs, linearity of oscillators, absence of coupling) is violated, then the estimators $\hat{\gamma}_{1,2}$ and $\hat{\delta}$ may become biased and the expressions for their confidence bands may no longer correspond to 95% reliability.

In this work, we vary different properties of oscillators and find out where the estimators $\hat{\gamma}_{1,2}$ and $\hat{\delta}$ are still reliable. To accomplish this, we aim at answering the following questions:

- "under what conditions the estimators $\hat{\gamma}_{1,2}$ remain unbiased?";
- "under what conditions the probability of erroneous conclusions about coupling presence and directionality remains less than 5 % ?";
- since one may also obtain indefinite conclusions about coupling character, i.e. that it is impossi-

ble to detect coupling presence or directionality with confidence, an important question is "under what conditions the probability of correct conclusions about coupling presence and directionality is high?". To be concrete, we determine when this probability is greater than 75 %.

To answer the first question we calculate biases of estimators $\hat{\gamma}_{1,2}$, which are equal by definition to ($E[\hat{\gamma}_{1,2}] - c_{1,2}$) where $E[\hat{\gamma}_i]$ is the expectation of $\hat{\gamma}_i$. We estimate $E[\hat{\gamma}_i]$ as the empiric mean value of $\hat{\gamma}_i$ over an ensemble of 1000 time series, standard error of the mean is regarded as the error in the obtained estimate of $E[\hat{\gamma}_i]$. If the true value of $c_i^2$ is not known *a priori*, it is estimated as the value of $\hat{\gamma}_i$ for a long time series with $N$ = 200 000. The estimator $\hat{\gamma}_i$ is regarded biased if the obtained estimate $\left|E[\hat{\gamma}_i] - c_i^2\right|$ is greater than double error in the estimate of $E[\hat{\gamma}_i]$.

To answer the second (third) question, we count the number of erroneous (correct) conclusions about coupling presence and directionality over the same ensemble of 1000 time series and check whether it is less than 5 % (greater than 75 %).

If not stated otherwise, time series of phases in numerical experiments are of the length $N$ = 1000. In Sec.3.2-3.4 they are generated by a system of stochastic differential equations using Euler integration technique with the step size $h = 0.01\pi$. Sampling interval $\Delta t$ may not coincide with $h$: we use $\Delta t = 20h$ and $\Delta t = h$ in Sec.3.1, $\Delta t = 20h$ in Sec.3.2-3.3, $\Delta t = 10h$ in Sec. 3.4. The value of $\tau$ is always taken to be $2\pi$ which is approximately equal to a basic period in all examples, i.e. a time interval over which the phase increases by $2\pi$. Following Refs. 1, 17, and 18, we use the third-order polynomials $F_i$. We calculate also mean phase coherence[15] $\rho = \left|\langle \exp\{j(\phi_2 - \phi_1)\}\rangle\right|$, where angle brackets stand for the time average, which quantifies the degree of synchrony in the systems' oscillations, to check whether it can always warn about inapplicability of the method. Such warning can be generated, at least sometimes, if $\rho$ >0.6 as observed in Ref. 23.

## 3. RESULTS

### 3.1. Influence of noise properties

The estimators $\hat{\gamma}_{1,2}$ and $\hat{\delta}$ are rigorously applicable if the noise terms $\varepsilon_{1,2}$ in (6) are Gaussian and their ACFs are linearly decreasing down to zero over the interval $[0, \tau]$. To check to what extent these conditions are necessary, we apply the method to estimate coupling from time realizations of the system (6) with different properties of $\varepsilon_{1,2}$. In the following, we fix $\omega_1 = 1.1$, $\omega_2 = 0.9$.

*Variation of the autocorrelation time.* Noises $\varepsilon_{1,2}$ are taken to be Gaussian with ACF linearly decreasing down to zero over the interval $[0, T]$. We call $T$ "the autocorrelation time" and vary it in the range $[0,10\tau]$. A noise realization for a necessary value of *T* is generated with the aid of moving average filter applied to the sequence of i.i.d. Gaussian random values. Sampling interval is taken to be $\Delta t = 0.2\pi$, i.e. $\tau = 10\Delta t$. Noise level $\sigma_1 = \sigma_2 = \sigma$ is varied in the range $[0,0.6]$.

As a result of calculations, we found that for all *T* and $\sigma$, the number of erroneous conclusions about coupling presence does not exceed 4 % and the estimators $\hat{\gamma}_{1,2}$ and $\hat{\delta}$ remain unbiased. E.g., for $\sigma = 0.12$ and $T = 10\tau$, we obtained $E[\hat{\gamma}_1] \approx 1.3 \cdot 10^{-4}$ with standard error of the mean $3.1 \cdot 10^{-3}$. Thus, as one can judge from this particular example, variation of the ACFs of the noises $\varepsilon_{1,2}$ does not itself bound applicability of the estimators $\hat{\gamma}_{1,2}$ and $\hat{\delta}$.

*Different probability distributions.* Next, we consider noises $\varepsilon_{1,2}$ with qualitatively different probability density functions (PDFs). ACFs remain the same as above, i.e. linearly decreasing down to zero over the interval $[0,\tau]$. To simplify calculations, we use sampling interval $\Delta t = \tau = 2\pi$ so that ACFs decreases down to zero over single sampling interval and one can generate noise realizations $\varepsilon_{1,2}(t_i)$ just as the sequence of i.i.d. random values[42]. We consider the following PDFs:

- unsmooth PDF – uniform distribution on a finite interval;
- asymmetric PDF – demeaned chi-square distribution with one degree of freedom;
- bimodal PDF – random alternation of values drawn from two Gaussian distributions with the same variance and different expectations.

Noise intensities $\sigma_1 = \sigma_2 = \sigma$ are varied in the range $[0,0.6]$. The results are practically the same for all PDFs and noise levels. Namely, the estimators are unbiased and the number of erroneous conclusions about coupling presence is less than 5 %. E.g., for the uniform distribution with $\sigma = 0.12$ we obtained $E[\hat{\gamma}_1] \approx 1.0 \cdot 10^{-4}$ and standard error of the mean $3.7 \cdot 10^{-3}$, the number of erroneous conclusions is 5%. For asymmetric PDF with $\sigma = 0.12$, we have $E[\hat{\gamma}_1] \approx 2.3 \cdot 10^{-4}$ and standard error of the mean $3.8 \cdot 10^{-3}$, the number of errors is 5%. For bimodal PDF with $\sigma = 0.12$, $E[\hat{\gamma}_1] \approx 1.4 \cdot 10^{-4}$ and standard error of the mean $3.6 \cdot 10^{-3}$, the number of errors is 4.9 %. So, the form of the PDFs does not seem to affect applicability of the estimators also.

This result can be understood intuitively based on the *robust estimation* ideas[43]. It is known from the linear regression theory that ordinary least-squares estimators of the regression coefficients are *statistically efficient* in the case of independent normally distributed observation errors. Variation of the distribution in a wide class (all distributions with variance less than a certain finite value) does not change significantly the accuracy of the estimators. Non-zero correlations between the ob-servation errors often do not affect it also. In our case, the estimators $\hat{\gamma}_{1,2}$ and $\hat{\delta}$ are based on the least-squares estimators of the model coefficients, so the robustness of the latter could carry over to $\hat{\gamma}_{1,2}$ and $\hat{\delta}$. A more serious problem arises if the distributions of $\varepsilon_{1,2}(t_i)$ strongly depend on the current phases $\phi_1(t_i), \phi_2(t_i)$. Such a case is encountered below for strongly nonlinear or strongly coupled oscillators.

3.2. Influence of the individual nonlinearities of oscillators

To check to what extent the properties of the estimators deteriorate when oscillators are nonlinear, we calculate $\hat{\gamma}_{1,2}$ and $\hat{\delta}$ from time realizations of the system (3) with $G_i(\phi_1,\phi_2) = \omega_i + b\cos\phi_i$, where $\omega_1 = 1.1$, $\omega_2 = 0.9$, and $\xi_{1,2}$ are Gaussian white noises. The coefficient $b$ determines the "phase nonlinearity strength". Noise level $\sigma_1 = \sigma_2 = \sigma$ is varied in the range $[0, 0.66]$, the value of $b$ in the range $[0, 0.80]$, $\Delta t = 0.2\pi$.

The results for $\hat{\gamma}_1$ are shown in Fig. 1 (a). They are analogous for $\hat{\gamma}_2$. The estimator $\hat{\gamma}_1$ is unbiased and the probability of erroneous conclusion about coupling presence is less than 5 % in the region to the left from the solid line, i.e. up to sufficiently strong nonlinearity $b$ = 0.3-0.7. The values of mean phase coherence $\rho$ are shown in Fig. 1 (b) with grayscale, $\rho$ increases with nonlinearity to some extent since distribution of the wrapped phase difference $\phi_2 - \phi_1$ on the interval $[0, 2\pi]$ becomes less uniform. However, $\rho$ is relatively small even to the right from the solid line in Fig. 1 (a) where $\hat{\gamma}_1$ is biased or error probability is high, so $\rho$ cannot reliably detect such situations here.

Let us express the result in "physical" units, i.e. considering contributions of the nonlinear term $b\cos\phi_1$ and noise term $\xi_1$ into the dynamics with respect to contribution of the term $\omega_1 = 1.1$, the latter can be interpreted as the influence of the linear component of the restoring force of the first oscillator. We express the relative value of nonlinearity as $b/\omega_1$. Relative noise level is $\sigma/\sqrt{2\pi\omega_1}$ which is derived as follows. Contribution of white noise $\xi_1$ over the period $T_1 = 2\pi/\omega_1$ is equal to $\sigma\sqrt{T_1}$ (it is a standard deviation of the integral of $\xi_1$ over time interval $T_1$) and contribution of the linear restoring force is $\omega_1 T_1 = 2\pi$. In the new relative units, the numerical values along the horizontal axes in Fig. 1 remain practically unchanged, while the values along the vertical axes decrease approximately 2.5 times. We conclude that the coupling estimators are unbiased and probability of erroneous conclusion about coupling presence is less than 5 % for noise intensity in the range 0-25 % of the linear component of the restoring force and phase nonlinearity strength up to 30-70%.

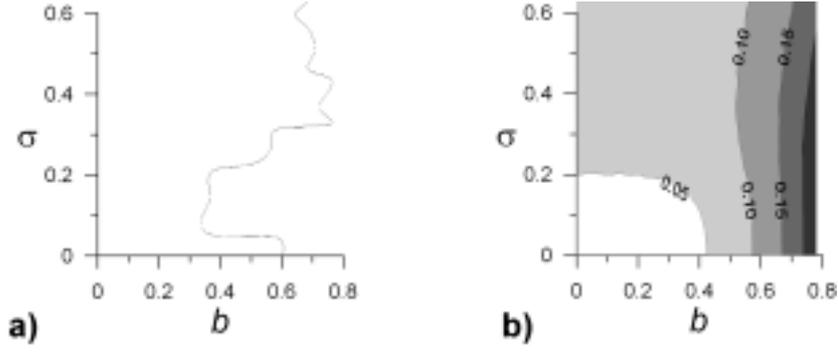

**FIG. 1.** a) Regions of the coupling estimators applicability on the plane "nonlinearity – noise" for uncoupled oscillators – system (3) with $G_i = \omega_i + b\cos\phi_i$. $\hat{\gamma}_1$ is unbiased and probability of erroneous conclusions about the influence 2→1 is less than 5% to the left from the solid line. b) Mean phase coherence values in grayscale, they are shown in the same manner in the figures 2-6 below, where pictures with the regions of applicability and mean phase coherence are combined together.

Similar results are observed for different nonlinearities. We present two additional examples here. The first one is the system (3) with $G_i(\phi_1,\phi_2) = \omega_i + b\cos 3\phi_i$. Fig. 2 (a) shows that the estimators are applicable up to nonlinearities of 40-80 % of the linear component of restoring force. The second example is the system (3) with $G_i = \omega_i + 0.76b(1-\varphi_i^2)\exp(-\varphi_i^2/2)$, where $\varphi_i \equiv (\phi_i \bmod 2\pi) - \pi$ and multiplier 0.76 provides root-mean-square value of the term $0.76(1-\varphi_i^2)\exp(-\varphi_i^2/2)$ over the interval $[-\pi, \pi]$ equal to 0.5 as for the trigonometric nonlinearities considered above. Fig. 2 (b) shows that here for noise levels up to 20% even stronger nonlinearity (up to 100-300%) is allowable.

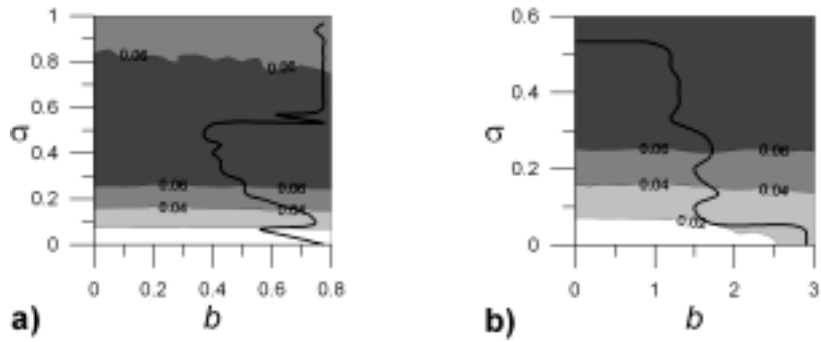

**FIG. 2.** Regions of the coupling estimators applicability (to the left from the solid line) on the plane "nonlinearity – noise" for uncoupled oscillators: a) system (3) with $G_i = \omega_i + b\cos 3\phi_i$; b) system (3) with $G_i = \omega_i + 0.76b(1-\varphi_i^2)\exp(-\varphi_i^2/2)$, $\varphi_i = (\phi_i \bmod 2\pi) - \pi$ .3.3. Influence of coupling strength.

Thus, the domain of the estimators' applicability appears quite significant with respect to the nonlinearity strength for all three cases considered, different nonlinearities manifesting themselves in a very similar manner. In other words, linearity of the oscillators is not a necessary condition for the estimators' applicability and can be moderately violated.

To check to what extent the estimators $\hat{\gamma}_{1,2}$ and $\hat{\delta}$ are applicable when considerable coupling between oscillators is present, we calculate them from time realizations of the system (3) with $\omega_1 = 1.1$, $\omega_2 = 0.9$, Gaussian white noises $\xi_{1,2}$, $\Delta t = 0.2\pi$, and different coupling functions. First, we consider $G_{1,2}(\phi_1, \phi_2) = \omega_{1,2} + k_{1,2} \sin(\phi_{2,1} - \phi_{1,2})$. The coefficients $k_1, k_2$ determine the coupling strengths. We consider the cases of unidirectional and bidirectional coupling in turn.

*Unidirectional coupling.* $k_1 = 0$, the value of $k_2 = k$ is varied in the range $(0, 0.25]$, noise level $\sigma_1 = \sigma_2 = \sigma$ in the range $[0, 0.5]$. In Fig. 3 (a) we show the "triangle" region where the estimates $\hat{\gamma}_{1,2}$ are unbiased (this condition determines the right boundary which is close to vertical straight line) and the number of correct conclusions about coupling strength is greater than 75% (this condition determines the curved left boundary which makes sense as a minimal reliably identifiable coupling strength for a given noise level). The estimators are erroneous if $\rho > 0.8$, see Fig. 3 (a). So, the "rule of thumb" that $\rho$ close to 0.6 is a sign of danger[23] for application of the EMA seems to be roughly confirmed here.

The causes of bias in the estimates in the case of large *k* are following: (i) synchronization for low noise levels [Fig. 3 (a)], (ii) nonlinearity of the phase dynamics induced by the presence of coupling for high noise levels. At a given noise level, the best situation is an intermediate strength of unidirectional coupling, since at weak coupling the probability of correct conclusion is low due to noise and at strong coupling the estimates become biased due to synchronization or just phase nonlinearity. Domain of the estimators' applicability widens with the time series length at fixed sampling frequency, see a big region in Fig. 3 (b) for time series length *N* = 4000. Note that right boundary is not a vertical line any more: for higher noise level stronger coupling is acceptable since intensive noise prevents synchronization that is good for the modified EMA application. For *shorter* time series strong noise is not so useful [almost vertical right boundary for *N* = 1000, Fig. 3 (a) and (b)] because there is no enough data to reliably extract information about coupling. If the time series length is increased only due to increase in sampling frequency, the results almost do not change, see the dashed line in Fig. 3 (b). The reason is that new data points sampled from the same time interval are highly correlated with the data already present, so that the former provide almost no new infor-mation about the dynamics. Thus, it is not reasonable to aim at a very high sampling frequency, it is enough to use a frequency sufficient for reliable phase extraction (20 data points per basic period[16]).

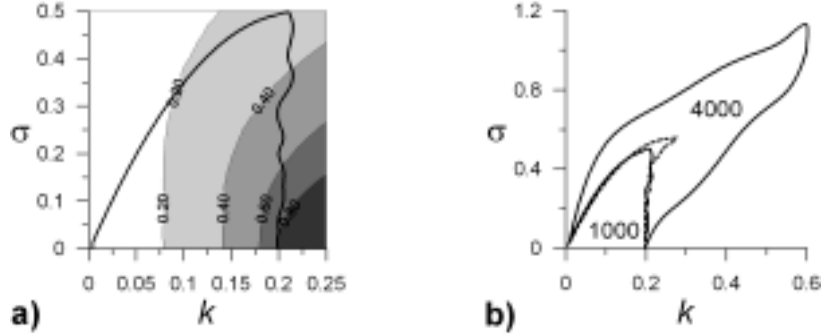

**FIG. 3.** Regions of the coupling estimators applicability on the plane "coupling – noise" for unidirectionally coupled phase oscillators – system (3) with $G_1 = \omega_1$, $G_2 = \omega_2 + k\sin(\phi_1 - \phi_2)$. a) A base case of $N=1000$ and $\Delta t = 0.2\pi$ (a time series comprises 100 basic periods). b) Comparison of different time series lengths. Small region is reproduced again for $N=1000$, $\Delta t = 0.2\pi$ (100 basic periods in a time series); dashed line bounds the applicability region for $N=4000$, $\Delta t = 0.05\pi$ (again 100 basic periods); large region is for $N=4000$, $\Delta t = 0.2\pi$ (400 basic periods).

In relative units ($k/\omega_2$ and $\sigma/\sqrt{2\pi\omega_2}$) one observes that the estimators work well for coupling strength up to 20% of the linear restoring force. Coupling strength of 20% can be identified reliably from a time series of the length $N = 1000$ for noise intensity up to 20%. Arbitrary weak coupling can be detected reliably if noise level is sufficiently low: the left boundary in Fig. 3 (a) is an almost straight line $\sigma \approx 4.6k$ for weak couplings.

Similar conclusions can be drawn for different coupling functions. Fig. 4 (a) shows the results for the system (3) with $G_1 = \omega_1$, $G_2(\phi_1, \phi_2) = \omega_2 + k\sin(3\phi_1)$. Here synchronization does not take place even for very strong coupling, so that no problem arises with increase in coupling strength. Only the boundary determined by high noise (low number of correct conclusions about coupling) is observed. Fig. 4 (b) is obtained for the system (3) with $G_2(\phi_1, \phi_2) = \omega_2 + 1.26k\Delta\phi \exp(-\Delta\phi^2/2)$, where $\Delta\phi$ is phase difference $\phi_1 - \phi_2$ wrapped to the interval [-$\pi$, $\pi$]. The region of the estimators' applicability is up to coupling strength of 40 %. Mean phase coherence reaches the value of 0.6 within this region, that is the value of $\rho \geq 0.6$ as an indicator of danger is well confirmed here.

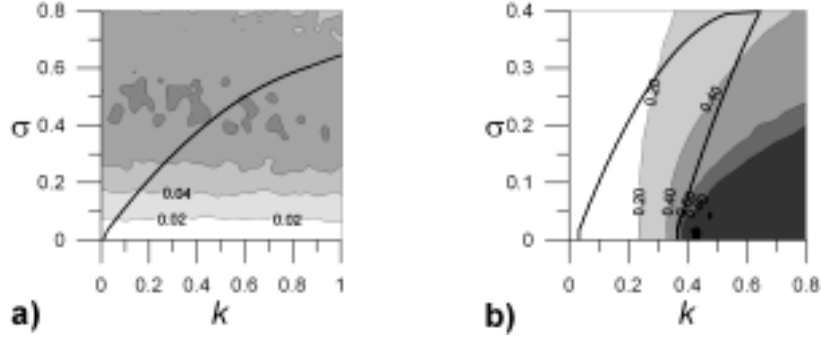

**FIG. 4.** Regions of the coupling estimators applicability on the plane "coupling – noise" for unidirectionally coupled phase oscillators: a) system (3) with $G_2 = \omega_2 + k\sin(3\phi_1)$ (region of applicability is below the thick line); b) system (3) with $G_2(\phi_1,\phi_2) = \omega_2 + 1.26k\Delta\phi\exp(-\Delta\phi^2/2)$, $\Delta\phi$ is the phase difference $\phi_1 - \phi_2$ wrapped to [-$\pi$, $\pi$].

Thus, the estimators are widely applicable in respect of coupling intensity in all examples.

*Bidirectional coupling.* $G_{1,2}(\phi_1,\phi_2) = \omega_{1,2} + k_{1,2}\sin(\phi_{2,1} - \phi_{1,2})$, $k_1 = k$, $k_2 = k + 0.02$. The value of $k$ is varied in the range [0,0.1]. The value of coupling asymmetry $k_2 - k_1$ is held constant. Noise level $\sigma_1 = \sigma_2 = \sigma$ is varied in the range [0,0.12].

The results of calculations are shown in Fig. 5. The region of the coupling estimators efficiency is bounded on the right (i.e. for large coupling strength). $\rho$ reaches a value of 0.8 within this region. Again, there are two causes that limit the estimators' applicability. For low noise level, there is mainly an increase in oscillations' synchrony, which induces biases in the estimators. For high noise level, there is a significant scattering of the estimates' values, which induces small probability of correct conclusions about coupling character.

In relative units ($k/\omega_2$ and $\sigma/\sqrt{2\pi\omega_2}$) the limits of applicability are up to 8% for coupling strength at noise level up to 2% and up to 2 % for coupling strength at noise level about 5%. Noise level of 5 % is the greatest allowable one. Thus, for bidirectional coupling the method also works properly for significant intervals of coupling strength and noise intensity values. But the region of applicability is narrower than that presented in Fig.3 (a) since asymmetry in coupling is small. Bounds of the region of applicability move apart with increase in coupling asymmetry and fixed overall coupling strength. So, the case of unidirectional coupling considered above is the easiest one for the determination of coupling directionality.

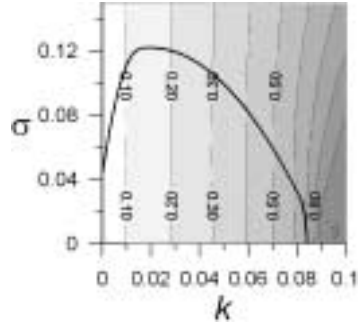

**FIG. 5.** Regions of the coupling estimators applicability on the plane "coupling – noise" for bidirectionally coupled phase oscillators – system (3) with $G_{1,2} = \omega_{1,2} + k_{1,2}\sin(\phi_{2,1} - \phi_{1,2})$.

3.4. Van der Pol and van der Pol – Duffing oscillators

More realistic is a situation where one observes not phases directly but rather some variables from which one needs to calculate phases and, hence, may introduce some additional errors. To simulate such a situation, first, we take coupled van der Pol oscillators as an object:

$$d^2 x_{1,2}/dt^2 = 0.2(1 - x_{1,2}^2)dx_{1,2}/dt - \omega_{1,2}^2 x_{1,2} + k_{1,2}(x_{2,1} - x_{1,2}) + \xi_{1,2}, \qquad (7)$$

where $\omega_{1,2}$ are angular frequencies, $\omega_1 = 1.02$, $\omega_2 = 0.98$, $\xi_{1,2}$ are Gaussian white noises. We consider both the usual case of independent sources of dynamical noise $\xi_{1,2}$ and common noise $\xi_1 = \xi_2$. We take the variables $x_1, x_2$ as observables, both the case of absence and presence of observational noise are considered. The signals are quasi-harmonic for moderate noise levels considered here, so the phases of oscillations are readily calculated with the aid of Hilbert transform even without filtering. Sampling interval is $\Delta t = 0.1\pi$ that corresponds to 20 data points per basic period. Original time series length is $N$ = 1400. After calculation of phases, we discard 200 values at each edge to avoid edge effects[16]. So, the resulting time series of each phase comprises 1000 data points, i.e. 50 basic periods. The oscillators possess individual phase nonlinearity. Noise in the phase dynamics equations is not precisely Gaussian and white. So, this object represents simultaneous violation of several conditions for the estimators applicability. We consider unidirectional coupling: $k_1 = 0$, the value of $k_2 = k$ is varied in the range $(0, 0.08]$.

*Independent sources of dynamical noise, observational noise is absent.* The level of dynamical noise $\sigma_1 = \sigma_2 = \sigma$ is varied in the range $[0, 0.16]$. In Fig. 6 (a) we present the region where the estimators are unbiased (right boundary) and the probability of correct conclusion about coupling presence is greater than 75 % (left boundary). $\rho$ reaches approximately 0.7 within the region, again in good agreement with the rule that $\rho > 0.6$ is dangerous for the method application. The boundaries are almost straight lines. As usually, $\rho$ becomes greater to the right from this region and estima-tors become biased due to significant of oscillations. Left boundary is determined by low probability of correct conclusions due to noise. The results are quite analogous to Fig. 3 (a) in Sec.3.3. In relative units, we express contributions of coupling and noise to the phase dynamics as $k\sqrt{\langle x_1^2 \rangle + \langle x_2^2 \rangle} / \omega_2^2 \sqrt{\langle x_2^2 \rangle}$ and $\sigma\sqrt{T_2} / \omega_2^2 \sqrt{\langle x_2^2 \rangle}$, respectively. Using the observed value $\langle x_{1,2}^2 \rangle \approx 2.3$ at weak coupling, we obtain relative contributions of coupling and noise approximately equal to $1.4k$ and $1.4\sigma$. Range of applicability here is up to 6.5 % for coupling strength [less than for the phase oscillators, Fig. 2 (a)] and up to 17 % for noise level (approximately the same as for the phase oscillators).

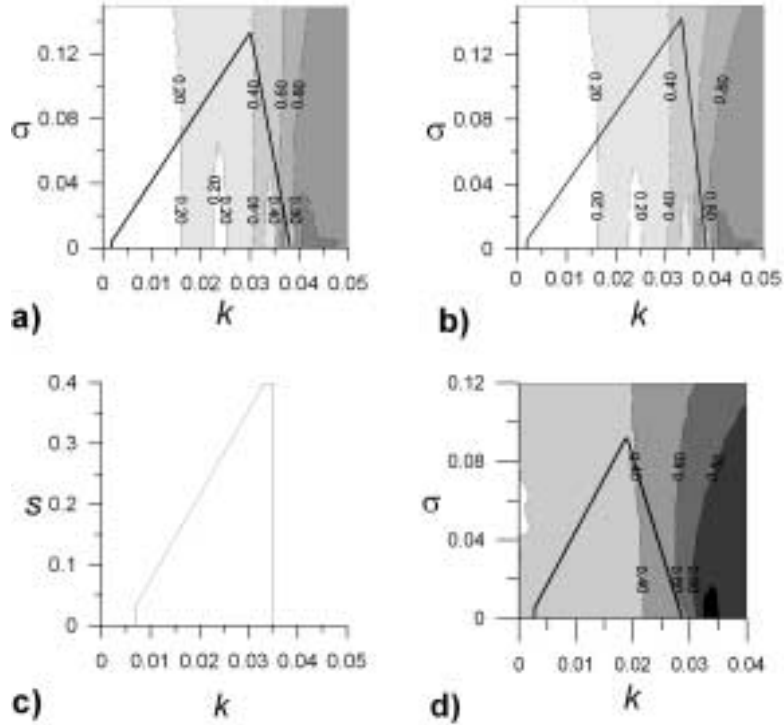

**FIG. 6.** Regions of the coupling estimators applicability on the plane "coupling – noise" for unidirectionally coupled van der Pol oscillators (a-c) and van der Pol – Duffing oscillators (d). a) Van der Pol oscillators with independent noise sources. b) Van der Pol oscillators with a common dynamical noise. c) The region on the plane "coupling – observational noise" Van der Pol oscillators with independent noise sources and dynamical noise level $\sigma = 0.025$. d) Van der Pol – Duffing oscillators with independent dynamical noise sources, an analogue to Fig. 6 (a).

*Common dynamical noise, observational noise is absent.* Next, we analyze what changes if dynamical noise is the same for both oscillators. This case deserves a special attention since common driving (even stochastic one) can lead to an increase in the degree of synchrony between the two oscillators. This is undesirable for application of the coupling estimators analyzed here, see, e.g.,

Ref. 23. But the results of numerical experiments [Fig. 6 (b)] are surprisingly almost indistinguishable from the case of independent noises. So, at least for the range of parameters considered here, common noise is not an obstacle for the use of the coupling estimators $\hat{\gamma}_{1,2}$ and $\delta$.

*Observational noise is present, independent sources of dynamical noise.* Dynamical noise level is fixed to be $\sigma = 0.025$. Independent observational noises are added to the variables $x_{1,2}$. They are Gaussian and white and have the same standard deviation $s$ which is varied in the range [0,0.4]. Range of the method efficiency is shown in Fig. 6 (c). In relative units along vertical axis (ratio of $s$ to the standard deviation of $x_2$ which is equal to 1.5), one obtains that observational noise level up to 25 % may be allowable. Again, the range of the method applicability is not infinitesimally small but rather significant.

*Van der Pol – Duffing oscillators.* Finally, we present the results for a bit different nonlinearity of oscillators – unidirectionally coupled van der Pol – Duffing oscillators

$$d^2 x_{1,2}/dt^2 = 0.2(1 - x_{1,2}^2)dx_{1,2}/dt - \omega_{1,2}^2 x_{1,2} - x_{1,2}^3 + k_{1,2}(x_{2,1} - x_{1,2}) + \xi_{1,2}, \tag{8}$$

with $\omega_1 = 1.05$, $\omega_2 = 0.95$, $k_1 = 0$, $\xi_{1,2}$ independent Gaussian white noises, observational noise is absent. Other conditions are the same as above. Fig. 6 (d) shows that the results are very close to that reported for van der Pol oscillators. So, wide applicability of the estimators is again confirmed.

3.5. Application to EEG data

The data were recorded from intracranial depth electrodes implanted in a patient with medically-refractory temporal lobe epilepsy as part of routine clinical investigations to determine candidacy for epilepsy surgery. The recordings included several left temporal neocortical → hippocampal seizures that occurred over the course of a long partial status epilepticus, see an example in Fig. 7 (a). Two channels were analyzed: the first channel situated in the left hippocampus, the second channel in the left temporal neocortex, where the "interictal" activity between seizures at the time was comprised of pseudoperiodic epileptiform discharges. Visual analysis of the interictal-ictal transitions (shown with vertical dashed lines) determined that the seizures all started first in the neocortex, with an independent seizure subsequently beginning at the ipsilateral hippocampus. We analyzed 4 recordings, but here we present the results for only one of them for the sake of brevity, simply as an illustration of application of the method to a nonstationary real-world system.

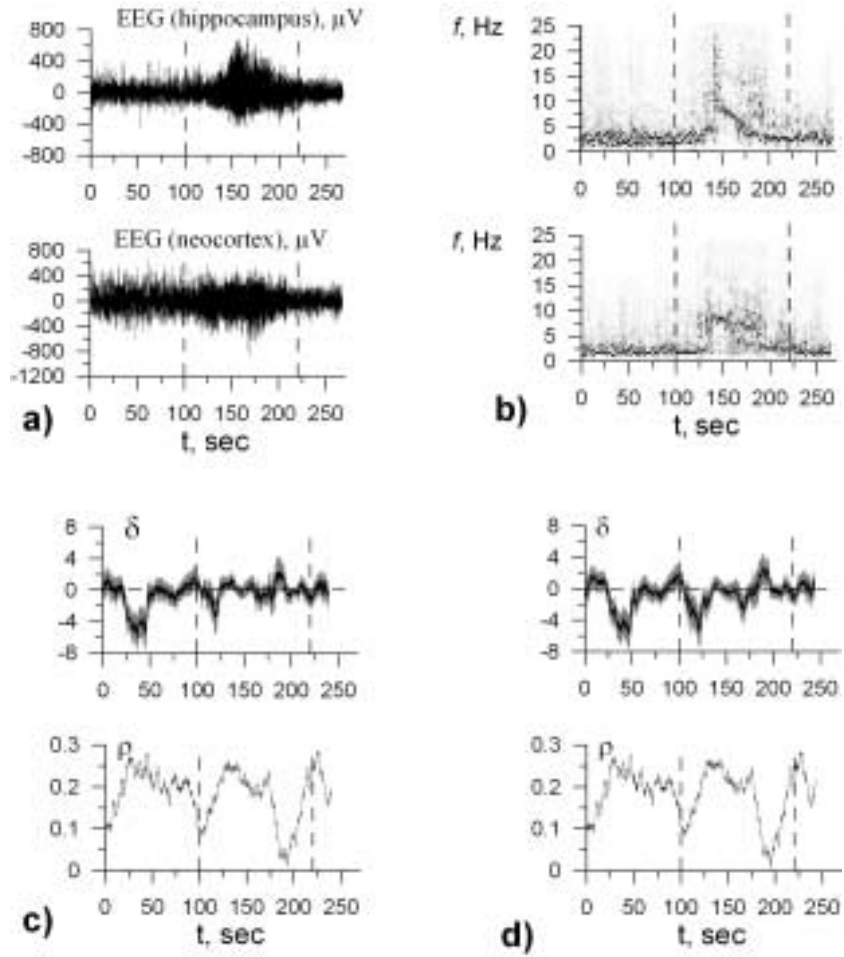

**FIG. 7.** a) EEG recordings from hippocampus (top) and neocortex (bottom). b) Spectrograms obtained with window length 512 data points with 3/4 overlap of the adjacent windows. c) Coupling directionality index and mean phase coherence for the phase obtained via Hilbert transform. Negative delta values correspond to coupling direction neocortex → hippocampus (approximately from the 20th second to the 50th second). Both signals are preliminarily low-pass filtered with cut-off frequency of 25 Hz. Window length is $N = 6000$, 500 phase values at each edge are discarded. d) Coupling directionality index and mean phase coherence for the phase obtained via wavelet transform, time scales correspond to the maxima of the scalograms, they are $s = 0.14$ sec for the hippocampal signal, and $s = 0.19$ sec for the neocortex signal. Window length is $N = 6000$, 130 phase values at each edge are discarded.

The time series of Fig. 7 (a) contains 4.5 min of depth electrode EEG (referential recording to scalp vertex electrode) recorded at a sampling frequency of 250 Hz. Their spectrograms are shown in Fig. 7 (b). One can see more or less significant peaks in power spectra for both channels. For the hippocampal channel: at frequency 3.2 Hz before the seizure (starting approximately at the 100th second and finishing approximately at the 220th second), 2.3 Hz after the seizure, and 7.1 Hz during the seizure. For the neocortex channel: at frequency 1.4 Hz before the seizure, 1.6 Hz after the seizure, and 7.1 Hz during the seizure.

We have computed coupling characteristics in a running window. The length of running window was changed from $N = 1000$ data points to $N = 10\,000$ data points. The value of $\tau$ was also changed from 25 to 1000 (the best results are expected for approximately $\tau = 33$ or 100 corresponding to the "main frequencies" of oscillations). The phases were determined using both versions of the analytic signal approach. For filtering with subsequent Hilbert transform, we tried different frequency bands: low-pass filter with cut-off frequencies 12.5 Hz and 25 Hz, band-pass filters with frequency bands around 2.5 Hz and 7.1 Hz, etc. For wavelet transform, we used Morlet wavelet with $\omega_0 = 2$ and different time scales $s$. In particular, we tried the time scales corresponding to the main peak of the scalogram for each signal which is $s = 0.14$ sec for the hippocampal signal, and $s = 0.19$ sec for the neocortex signal.

We present only one set of results, in Fig. 7 (c) and 7 (d) (gray tail denotes 95% confidence bands for $\delta$) obtained for windows of the length $N = 6000$. Coupling is regarded as significant if the confidence band does not include zero, e.g., gray tail does not intersect the abscissa axis. The preliminary results seem promising for localization of the epileptic focus, because a long interval (30 second length for the example shown) of significant predominant coupling direction neocortex → hippocampus is observed before the seizure. It can be considered as indication that epileptic focus is located near neocortex that agrees with a priori clinical information. Despite only a single example is presented, we note that the results are sufficiently robust and are observed for a significant range of values of the above mentioned window lengths and parameters of filters and wavelets.

Similar results of this type are observed for the three of the four analyzed recordings and not observed for one of them. Right now, we do not draw any definite conclusions about applicability of the method to localize epileptic focus. This is only the first attempt and, of course, more EEG recordings should be processed to quantify the method's *sensitivity* and *specificity*. This is a subject of ongoing research. Therefore, the results presented in this Section should not be overestimated, being rather an illustration of the way how to apply the method in practice and what kind of information one can expect from it.

## 4. DISCUSSION

Numerical experiments demonstrate that the estimators of coupling between oscillatory systems based on phase dynamics modeling are sufficiently widely applicable. Although they are derived under the strict assumption of linear uncoupled oscillators and independent sources of Gaussian white noise, they are valid for various dynamical noise properties including the case of common noise and finite (not negligibly small) strengths of nonlinearity, coupling, and observational noise. Thus, we conclude that

- variation of ACFs and PDFs of dynamical noise in difference equation for the phase dynamics does not affect applicability of the estimators;
- significant individual phase nonlinearity of the oscillators (up to 30-300 % of the linear component of the restoring force), unidirectional coupling strength (up to 30-40%), and observational noise (up to 25 %) may be allowable;
- the "rule of thumb" that mean phase coherence close to 0.6 warns about problems is generally confirmed, but strictly speaking any value of $0 < \rho < 0.8$ is neither a sufficient nor a necessary indicator for the estimators' applicability (on the one hand, they may be biased already for $\rho \approx 0.1$, on the other hand, they may be quite efficient even for $\rho \approx 0.8$);
- the probability of correct conclusion about coupling character is very small for weak coupling and large noise, but the corresponding bound moves apart with increase in time series length $N$ at fixed sampling frequency (increase in the number of basic periods contained in a time series);
- the probability of erroneous conclusion about coupling character is high for strong coupling due to considerable synchrony of oscillations, the corresponding bound depends relatively slightly on $N$ especially for low dynamical noise level.

When the modified EMA approach is not reliable (ill-defined phases, too strong noise, too strong coupling, too strong phase nonlinearity), other techniques may be efficient. Thus, well-known cross-correlation function and Fourier coherence are the methods of choice for very strong coupling and high level of observational noise. Moreover, they can be easily applied to short time series. However, the reverse of the medal is that they are capable of detecting only very strong and simple (linear) relationships between the oscillators' dynamics and, generally speaking, they may reveal only the *presence* of coupling, not *directionality*. There exist nonlinear generalizations of these techniques such as information-theoretic approaches[4,5] and nearest neighbors statistics in reconstructed state spaces[6,7,9]. These nonlinear techniques are more advanced in that they can reveal weak and complicated (nonlinear) interactions and their directionality, but simultaneously they are much more demanding in respect of time series length. Detailed comparison of one of the state space approaches and the modified EMA is given in Ref. 21. A strong coupling making the systems close to some type of synchronous regime (possibly nonlinear) is readily detected even in the presence of observational noise with "multidimensional phase coupling"[13] (for detection of interrelations in reconstructed state spaces) or mean phase coherence[15] (for detection of interrelations between the phases).

It is not possible here to discuss in detail various relationships between different coupling characterization techniques. However, we would like to stress that methods based on phase dynamics analysis seem to be the best for coupling characterization between weakly coupled oscillatory systems with well-defined phases, a situation widely spread in practice. The EMA is one of the best among these techniques as shown in Ref. 18. An approach very similar to the EMA was proposed by Kiemel et al in Refs. 14 and 20. It is based on construction of an empiric model for the phase dynamics in the form (3) with $G_{1,2}(\phi_1, \phi_2) = \omega_{1,2} + a_{1,2} \sin(\phi_{2,1} - \phi_{1,2})$. This approach is very close to the EMA with small $\tau$ since in this case the difference equation (4) is an accurate integration scheme for the differential equation (3). Analogously to the EMA, the approach of Kiemel et al would suffer from the bias problem for short time series and high dynamical noise level, but this bias could be corrected using considerations similar to that of Ref. 1. The main differences between the EMA and the approach of Kiemel et al are as follows. The latter does not require very weak coupling or nonlinearity, since model parameters are estimated via "honest" maximum likelihood method involving integration of the Fokker-Planck equation. Due to the complexity of calculations, it is very time consuming. The former is more demanding with respect to weakness of oscillators' nonlinearity and coupling (though not dramatically, as we showed here), but is much simpler and faster. Besides, the EMA is used with an optimal value of $\tau$, typically about a basic period of oscillations. Such a choice, as a rule, provides characteristics with significantly greater sensitivity to weak coupling than small $\tau$ which is close to the approach of Kiemel et al. Finally, both approaches can be regarded as slightly different versions of the same phase dynamics modeling approach.

The modified EMA analyzed here is the extension of the EMA to short time series so that it seems to be a very powerful method and deserves special attention. Based on considering several exemplary oscillators, we formulated empiric conditions for applicability of the corresponding coupling estimators. Even though these conditions could be somewhat different for other types of nonlinearity and coupling between oscillators, our results seem sufficiently representative and already allow to state that such conditions are rather mild. Thereby, we confirm the potential for the application of the estimators in practice to analyze real-world complex systems. In particular, our first attempt to apply them for epileptic focus localization from multichannel intracranial EEG recordings illustrated in the present paper looks promising.

## ACKNOWLEDGEMENTS

D.S. and B.B. are grateful to A. Pikovsky and M. Rosenblum for useful comments and advice. This work is supported by the RFBR (grant No. 05-02-16305), program BRHE (REC-006), President of Russia (grant No. MK-1067.2004.2), and Russian Science Support Foundation. J.L.P.V. ac-knowledges the support of the Natural Sciences and Engineering Research Council of Canada (NSERC) through a Discovery grant.

APPENDIX: EXPRESSIONS FOR COUPLING ESTIMATORS

Formulas for estimators $\hat{\gamma}_{1,2}$ are derived for linear uncoupled oscillators under influence of Gaussian white noise[1]. They are expressed in terms of estimators of coefficients of the model (4), where functions $F_i$ are trigonometric polynomials

$$F_i = \sum_{m,n}[a_{i,m,n}\cos(m\phi_1 + n\phi_2) + b_{i,m,n}\sin(m\phi_1 + n\phi_2)], \quad i=1,2. \quad (A1)$$

Coefficient estimates $\hat{a}_{i,m,n}$ and $\hat{b}_{i,m,n}$ are obtained via the least-squares routine and estimates of their variances are

$$\hat{\sigma}^2_{\hat{a}_{i,m,n}} = \frac{2\hat{\sigma}^2_{\varepsilon_i}}{N-k}\left\{1 + 2\sum_{l=1}^{k-1}\left(1-\frac{l}{k}\right)\cos\left[\frac{l(m\hat{a}_{1,0,0} + n\hat{a}_{2,0,0})}{k}\right]\exp\left[-\frac{l(m^2\hat{\sigma}^2_{\varepsilon_1} + n^2\hat{\sigma}^2_{\varepsilon_2})}{2k}\right]\right\}, \quad (A2)$$

where $\hat{\sigma}^2_{\varepsilon_i}$ is the estimate of variance of the noise $\varepsilon_i$ in difference equations (4) which reads

$$\hat{\sigma}^2_{\varepsilon_i} = \frac{1}{N-k-L_i}\sum_{j=1}^{N-k}\left[\Delta_i(t_j) - \frac{1}{N-k}\sum_{l=1}^{N-k}\Delta_i(t_l)\right]^2, \quad (A3)$$

where $L_i$ is the number of coefficients of the polynomial $F_i$. Expression for the variances of $\hat{b}_{i,m,n}$ is the same. Estimator $\hat{\gamma}_1$ is expressed via estimates of coefficients and their variances as

$$\hat{\gamma}_1 = \sum_{m,n} n^2\left(\hat{a}^2_{1,m,n} + \hat{b}^2_{1,m,n} - 2\hat{\sigma}^2_{\hat{a}_{1,m,n}}\right). \quad (A4)$$

Expression for $\hat{\gamma}_2$ is analogous. Directionality index is defined as $\hat{\delta} = \hat{\gamma}_2 - \hat{\gamma}_1$.

Estimate of the variance of $\hat{\gamma}_1$ reads

$$\hat{\sigma}^2_{\hat{\gamma}_1} = \begin{cases} \sum_{m,n} n^4(\hat{\sigma}^2_{\hat{a}^2_{1,m,n}} + \hat{\sigma}^2_{\hat{b}^2_{1,m,n}}), & \hat{\gamma}_1 \geq 5\sqrt{\sum_{m,n} n^4(\hat{\sigma}^2_{\hat{a}^2_{1,m,n}} + \hat{\sigma}^2_{\hat{b}^2_{1,m,n}})}, \\ \frac{1}{2}\sum_{m,n} n^4(\hat{\sigma}^2_{\hat{a}^2_{1,m,n}} + \hat{\sigma}^2_{\hat{b}^2_{1,m,n}}), & \end{cases} \quad (A5)$$

where

$$\hat{\sigma}^2_{\hat{a}^2_{i,m,n}} = \begin{cases} 2\hat{\sigma}^4_{\hat{a}_{i,m,n}} + 4(\hat{a}^2_{i,m,n} - \hat{\sigma}^2_{\hat{a}_{i,m,n}})\hat{\sigma}^2_{\hat{a}_{i,m,n}}, & \hat{a}^2_{i,m,n} - \hat{\sigma}^2_{\hat{a}_{i,m,n}} \geq 0, \\ 2\hat{\sigma}^4_{\hat{a}_{i,m,n}}, & \end{cases} \quad (A6)$$

and everything is the same for $\hat{\sigma}^2_{\hat{b}^2_{i,m,n}}$. Estimate of the variance of $\hat{\gamma}_2$ is derived analogously. For directionality index, one has the variance estimate $\hat{\sigma}^2_{\hat{\delta}} = \hat{\sigma}^2_{\hat{\gamma}_1} + \hat{\sigma}^2_{\hat{\gamma}_2}$. Confidence bands for the cou-pling estimates are expressed via their variances. Thus, 95% confidence bands were found semi-empirically: $[\hat{\gamma}_i - 1.6\hat{\sigma}_{\hat{\gamma}_i}, \hat{\gamma}_i + 1.8\hat{\sigma}_{\hat{\gamma}_i}]$ for $\hat{\gamma}_i$ and $[\hat{\delta} - 1.6\hat{\sigma}_{\hat{\delta}}, \hat{\delta} + 1.6\hat{\sigma}_{\hat{\delta}}]$ for directionality index.